\documentclass[prl,10pt,twocolumn,showpacs,amsmath,amssymb,floatfix,nobibnotes,superscriptaddress]{revtex4-2}

\usepackage{graphicx}
\usepackage{dcolumn}
\usepackage{bm}
\usepackage{color}
\usepackage{txfonts}
\usepackage{microtype}
\usepackage{hyperref}
\usepackage[english]{babel}
\usepackage{slashed}
\usepackage{gensymb}
\usepackage{epsfig}
\usepackage[normalem]{ulem}
\usepackage{amsmath}
\usepackage{array}
\usepackage{booktabs}
\allowdisplaybreaks[4]

\addto\captionsenglish{}
\begin{document}
\title{Generation of Relativistic Structured Spin-Polarized Lepton Beams}

\author{Zhong-Peng Li}
\affiliation{Ministry of Education Key Laboratory for Nonequilibrium Synthesis and Modulation of Condensed Matter, State key laboratory of electrical insulation and power equipment, Shaanxi Province Key Laboratory of Quantum Information and Quantum Optoelectronic Devices, School of Physics, Xi’an Jiaotong University, Xi’an 710049, China}

\author{Yu Wang}
\affiliation{Ministry of Education Key Laboratory for Nonequilibrium Synthesis and Modulation of Condensed Matter, State key laboratory of electrical insulation and power equipment, Shaanxi Province Key Laboratory of Quantum Information and Quantum Optoelectronic Devices, School of Physics, Xi’an Jiaotong University, Xi’an 710049, China}

\author{Yousef I. Salamin}
\affiliation{Department of Physics, American University of Sharjah, Sharjah, POB 26666 Sharjah,  United Arab Emirates}

\author{Mamutjan Ababekri}
\affiliation{Ministry of Education Key Laboratory for Nonequilibrium Synthesis and Modulation of Condensed Matter, State key laboratory of electrical insulation and power equipment, Shaanxi Province Key Laboratory of Quantum Information and Quantum Optoelectronic Devices, School of Physics, Xi’an Jiaotong University, Xi’an 710049, China}

\author{Feng Wan}
\affiliation{Ministry of Education Key Laboratory for Nonequilibrium Synthesis and Modulation of Condensed Matter, State key laboratory of electrical insulation and power equipment, Shaanxi Province Key Laboratory of Quantum Information and Quantum Optoelectronic Devices, School of Physics, Xi’an Jiaotong University, Xi’an 710049, China}

\author{Qian Zhao}
\affiliation{Ministry of Education Key Laboratory for Nonequilibrium Synthesis and Modulation of Condensed Matter, State key laboratory of electrical insulation and power equipment, Shaanxi Province Key Laboratory of Quantum Information and Quantum Optoelectronic Devices, School of Physics, Xi’an Jiaotong University, Xi’an 710049, China}

\author{Kun Xue}
\affiliation{Ministry of Education Key Laboratory for Nonequilibrium Synthesis and Modulation of Condensed Matter, State key laboratory of electrical insulation and power equipment, Shaanxi Province Key Laboratory of Quantum Information and Quantum Optoelectronic Devices, School of Physics, Xi’an Jiaotong University, Xi’an 710049, China}

\author{Ye Tian}
\affiliation{State Key Laboratory of High Field Laser Physics and CAS Center for Excellence in Ultra-intense Laser Science, Shanghai Institute of Optics and Fine Mechanics, Chinese Academy of Sciences, Shanghai, People’s Republic of China}
\affiliation{Center of Materials Science and Optoelectronics Engineering, University of Chinese Academy of Sciences, Beijing, People’s Republic of China}

\author{Jian-Xing Li}\email[Contact author: ]{jianxing@xjtu.edu.cn}
\affiliation{Ministry of Education Key Laboratory for Nonequilibrium Synthesis and Modulation of Condensed Matter, State key laboratory of electrical insulation and power equipment, Shaanxi Province Key Laboratory of Quantum Information and Quantum Optoelectronic Devices, School of Physics, Xi’an Jiaotong University, Xi’an 710049, China}
\affiliation{Department of Nuclear Physics, China Institute of Atomic Energy, P. O. Box 275(7), Beijing 102413, China}

\date{\today}
\begin{abstract}
	
Relativistic structured spin-polarized (SSP) particle beams, characterized by polarization structures, are of critical importance in a wide range of applications, such as material properties investigation, imaging, and information storage. However, generation of relativistic SSP beams faces significant challenges. Here, we put forward a novel method for generating relativistic SSP lepton beams via employing a moderate-intensity terahertz (THz) wave. Building upon our foundational work on velocity-matched spin rotation in dielectric-lined waveguides [Phys. Rev. Lett. 134, 075001 (2025)], we present the first demonstration of spin-polarization mode matching - a novel mechanism that establishes a direct relation between waveguide modes and beam polarization states. This breakthrough enables precise spatial control over spin structures at relativistic energies, generating customizable spin-polarization configurations such as spider-like, azimuthal, and helical structures, etc. Such SSP beams have the potential to generate high-energy structured photon beams and open a new avenue for research on relativistic structured particle beams, especially in nuclear physics, high-energy physics, materials science and atomic physics.
	
\end{abstract}

\maketitle

Structured particle beams, including polarization and phase structure, have shown significant applications across various fields. In particular, structured light beams \cite{Carmelo18JO,Halina17JO,Forbes21NPho}, which possess specific photon polarization patterns (e.g., radially and azimuthally polarized beams) and characteristic spatial phase structures (e.g., vortex beams), have found significant applications in various domains, including high-resolution microscopy \cite{Chen13OL, Stefan07S}, optical trapping \cite{Ashkin70PRL, Kozawa10OE, Otte20APR}, metrology \cite{Belmonte15Optica, Tang10PRL, Tang11S}, optical storage \cite{Fang20NPho}, and optical communication \cite{Wang12NPho, Barreiro08NPhys, Fang20NPho}, etc. In addition, vortex beams such as electrons \cite{Lloyd17RMP, Uchida10N, Shiloh15PRL, Harris15NPhy, Bliokh17PR} and neutrons \cite{Clark15N} have demonstrated significant importance in several research areas, including electron microscopy \cite{McMorran11S}, nanoparticle control \cite{Verbeeck13AM}, chirality \cite{Schattschneider14U, Rusz14PRB} and magnetism \cite{Verbeeck10N} of materials. However, spin-polarized (SP) particles have been spatially structured only within non-relativistic spintronic systems \cite{Koralek09N, Yumoto22SA, Kunihashi16NC,Li21N}, creating configurations like persistent spin helices \cite{Bernevig06PRL,Walser12NPhy} and Skyrmions \cite{Fu16PRL, Reichhardt22RMP, 24ShenNPho}.

Relativistic structured spin-polarized (SSP) lepton beams may have the ability to resolve spin spatially, effectively adding an extra degree of freedom, and find important applications in spin-dependent interactions such as polarized deep inelastic scattering \cite{Hughes99ARNPS, Bluemlein13PPNP, Anselmino95PR, Abelev09PRD, Adamczyk14PRL}. Moreover, SSP lepton beams could play an indirect role in chiral-selective chemistry \cite{Rosenberg11TCCS}. For instance, one can investigate the chiral effects of a SSP lepton beam on an initially racemic mixture (equal quantities of both enantiomers) to test the Vester-Ulbricht Hypothesis~\cite{Bonner91OLEB} and investigate the chirality issues of amino acids, carbohydrates and nucleic acids \cite{kessler85polarized}. In addition, SSP lepton beams may significantly contribute to the generation and control of high-brilliance structured light in the X- and $\gamma$-ray domains, both of which remain challenging today. Brilliant structured polarized $\gamma$-rays with degrees of polarization close to 95\% can be generated by nonlinear Compton scattering utilizing SSP lepton beams \cite{Li20PRL}. Once available, structured $\gamma$-ray beams can potentially be employed in many future trans-sectoral areas, such as particle \cite{Zhao23PRD,Kostenko18AJ,Adam21PRL} and photonuclear physics \cite{Weller09PPNP,Zilges22PPNP,Pietralla01PRL}. 

Although it is possible to generate transversely or longitudinally SP particle beams via Sokolov-Ternov effect~\cite{Sokolov66} or Bethe-Heitler process \cite{Abbott16PRL}, the generation of SSP particle beams remains a tremendous challenge. This difficulty arises because spin-manipulation devices, such as spin rotators~\cite{Moffeit06}, Wien filters \cite{Karimi12PRL} and Siberian snakes \cite{Derbenev78PA}, rely on dipole magnets and solenoids, resulting in high maintenance, high design costs, and limited spatial tunability. Moreover, the rapid development of ultraintense ultrafast laser techniques \cite{Mourou85OC, Yoon21OPtica, Kawanaka16JPCS, Cartlidge18S, Danson19HPLSE, Bahk04OL, Tiwari19OL, Pirozhkov17OE, Guo18OE, Zou15HPLSE, Gales18RPP, Bromage19HPLSE, Yaron18S} has opened new avenues for generating ultrafast polarized beams \cite{Xue23PRL, Piazza12RMP, Fedotov23PR, Song22PRL, Salamin06PR, Dai22MRE, Xue22FR, Wan20PLB, LYF19PRL, Seipt19PRA, CYY19PRL, Zhao22PRD,Vranic18SR, Xie17MRE}. Nevertheless, the transverse laser field substantially enhances both divergence angle and energy spread of the beam.

\begin{figure*}[t]
	\vspace{6pt}
	\centering\includegraphics[width=1\linewidth]{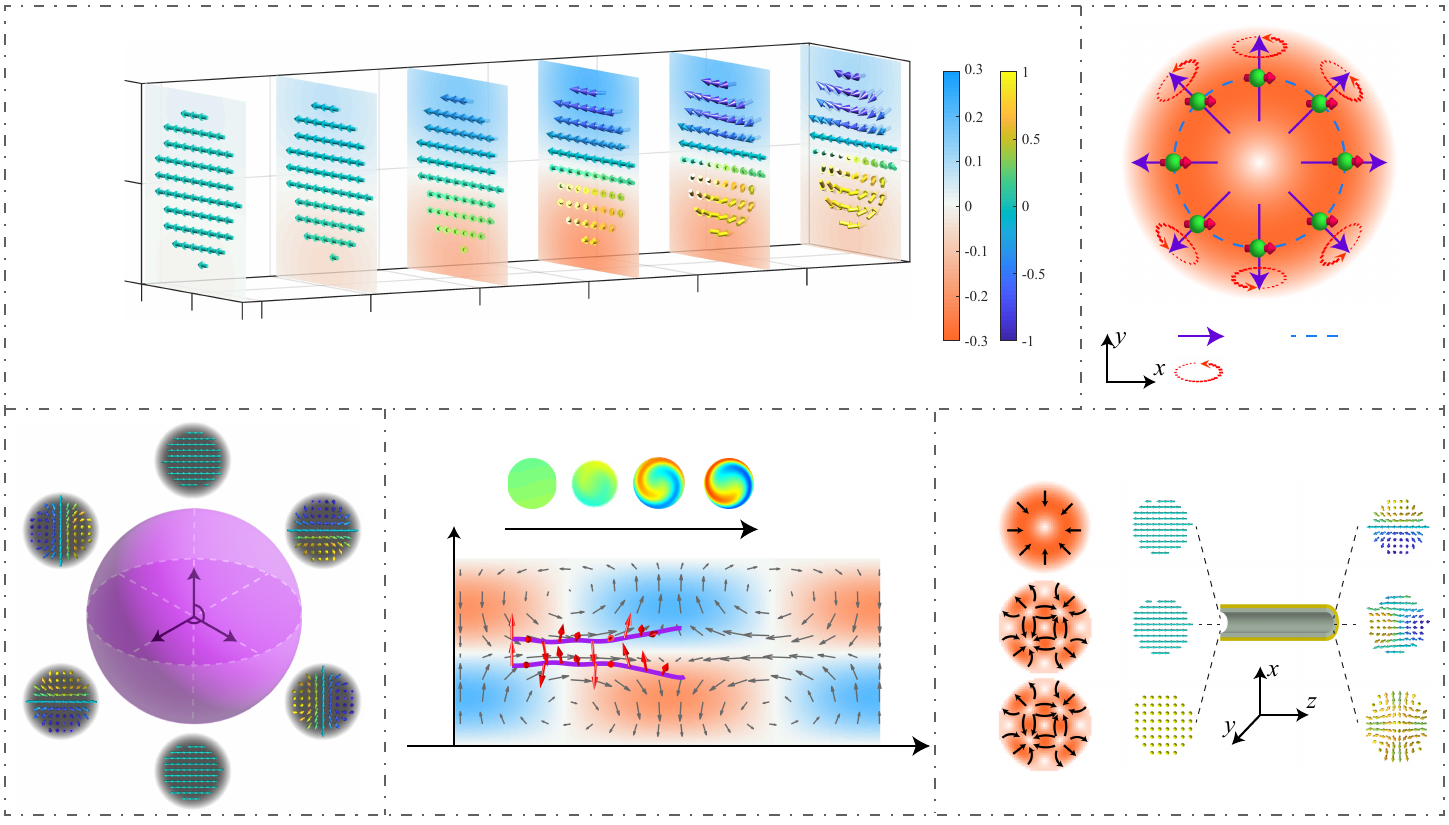}
	\begin{picture}(300,0)
		\put(-100,290){(a)}
		\put(-85 ,230){\rotatebox{90}{$y(\lambda_0)$}}
		\put(-75,270){0.4}
		\put(-68,235){0}
		\put(-78,200){-0.4}
		\put(-50,172){$x(\lambda_0)$}
		\put(-65,185){0.4}
		\put(-41,181){0}
		\put(-33,178){-0.4}
		\put(85,173){$z(\lambda_0)$}
		\put(-15,178){-1.5}
		\put(25,180){0}
		\put(60,182){1.5}
		\put(102,184){3}
		\put(138,186){4.5}
		\put(180,188){6}
		\put(226,282){$\mathrm{B_y}$}
		\put(248,282){$s_z$}
		\put(-41,275){$0T_0$}
		\put(5,278){$2T_0$}
		\put(50,281){$4T_0$}
		\put(95,284){$6T_0$}
		\put(141,286){$8T_0$}
		\put(187,286){$10T_0$}
		
		\put(280,290){(b)}
		\put(330,180){B}
		\put(372,180){E}
		\put(330,168){precession direction}
		
		\put(-100,146){(c)}
		\put(-38,72){$\psi_1$}
		\put(-30,85){$\psi_2$}
		
		\put(35,146){(d)}
		\put(78,145){SP structure evolution}
		\put(165,113){$\varphi_f$} \put(64,113){$\varphi_i$}
		\put(38,62){\rotatebox{90}{$x(\lambda_0)$}}	
		\put(44,89){0.5} \put(50,67){0} \put(40,45){-0.5} 
		\put(55,28){$\rm \pi$} \put(85,28){$\rm 0.5\pi$} \put(130,28){$\rm 0$}
		\put(161,28){$\rm -0.5\pi$} \put(198,28){$\rm -\pi$} 
		\put(130,19){$\varphi$}

		\put(230,146){(e)}
		\put(230,112){$\rm TE_{01}$}
		\put(230,78){$\rm TE_{22}$}
		\put(230,44){$\rm TE_{22}$}
		\put(253,135){mode}
		\put(295,132){input}
		\put(378,132){output}
		\put(336,90){DLW}

	\end{picture}
	\setlength{\abovecaptionskip}{-0.4 cm}
	\caption{Scenarios of generating a SSP lepton beam utilizing a THz wave in a DLW setup. (a) Evolution of the lepton spins when a TSP lepton beam propagates in the DLW with $\rm TE_{01}$ mode inside. Each slice represents the distribution of electron spins and intensity of magnetic field along the $y$-axis at different times. $\lambda_0$ denotes the wavelength of the THz wave, while $T_0$ represents its period. (b) Transverse sketch of spin precession of the leptons as they traverse along the azimuthal angle. (c) Poincar\'{e} sphere representation of the SP structures when the THz wave in DLW is $\mathrm {TE_{01}}$ mode. $\psi_1$ represents the angle between $x$-axis and initial direction of spin polarization of the beams, and $\psi_2$ denotes the angle between the maximum longitudinal polarization and the transverse plane. (d) Evolution in the $xz-$plane of the lepton spins with respect to THz phase in a $\rm TE_{01}$ mode. Evolution of the SP structure of the lepton beam is shown above the phase diagram. Purple lines represent lepton trajectories in the experienced phase, while red arrows represent lepton spin vectors. (e) The SP structures generated by different spin-polarization mode matching. }
	\label{fig1}
\end{figure*}
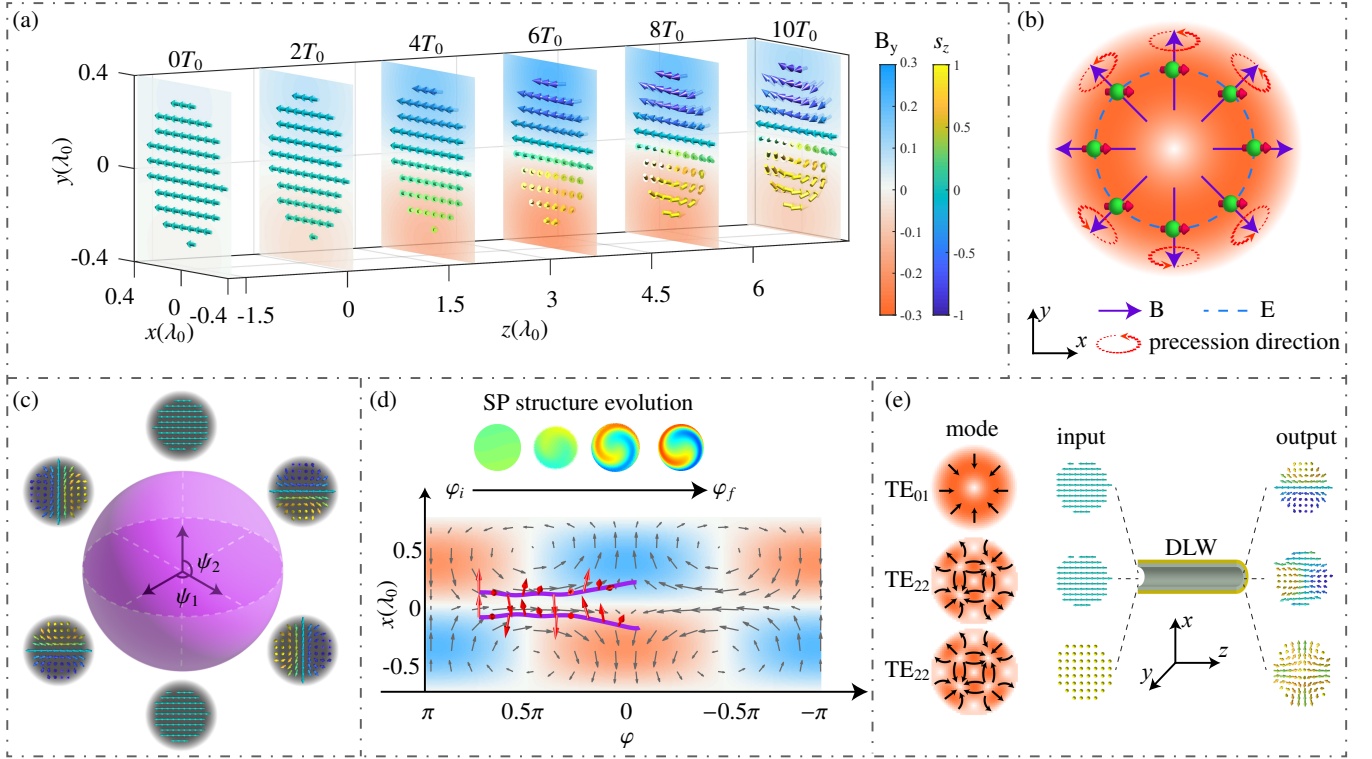

Over the past decade, terahertz (THz) waves have emerged as a promising tool for controlling electrons \cite{Snively20PRL, Ying24NPho, Zhang19Opti}. The flexibility in adjusting electromagnetic (EM) modes within a dielectric-lined waveguide (DLW) \cite{Nanni15NV, Xu21NPho, Fisher22NPho, ZDF20PRX, Tang21PRL, YXQ23NPho, Pacey19PRAB} further enhances this potential, enabling generation of SSP particle beams. Compared to radio-frequency technology, THz systems offer significant reductions in size and cost, along with increased breakdown thresholds \cite{Thompson08PRL, BD16NC}. In addition, when compared to infrared laser-driven dielectric microstructures \cite{Peralta13N, Breuer13PRL, England14RMP, Freemire23PRAB}, a DLW driven by a THz wave offers distinct advantages, such as accessible fabrication and high beam charge \cite{Hibberd20NPho}. Furthermore, the EM field inside a DLW exhibits greater stability and controllability compared to that in plasma wakefields \cite{Tajima79PRL}, making it a promising candidate for particle beam manipulation. It has been demonstrated that using THz waves in a DLW enables spin rotation of lepton beams \cite{my25PRL}. Since spin is fundamentally a quantum property described by the wavefunction, the ability of EM fields to manipulate lepton spins indicates their capacity to interact with wavefunctions. This interaction suggests the potential for precise control over wavefunction phases, which could enable generation of relativistic vortex beams.

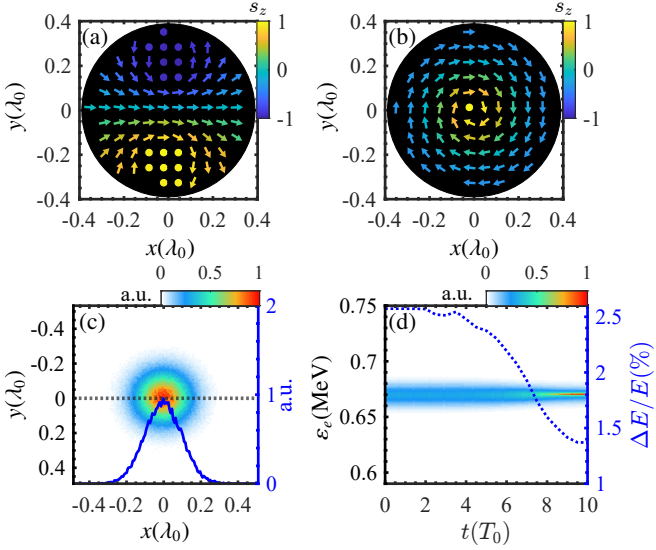
\begin{figure}[!t]
	\centering
    \flushright
		\includegraphics[width=0.96\linewidth]{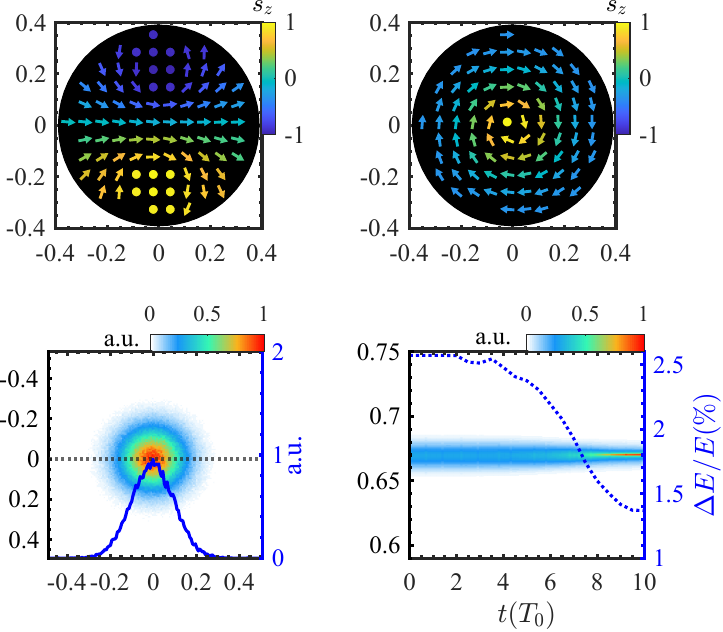}
		
		\begin{picture}(300,0) 
			\put(29,201){(a)}
			\put(0,168){\rotatebox{90}{$y (\lambda_0)$}}
			\put(53,121){$x (\lambda_0)$}
			
			\put(145,201){(b)}
			\put(118,168){\rotatebox{90}{$y (\lambda_0)$}}
			\put(168,121){$x (\lambda_0)$}
			
			\put(28,93){(c)}
			\put(0,59){\rotatebox{90}{$y (\lambda_0)$}}
			\put(51,14){$x (\lambda_0)$}
			
			\put(145,93){(d)}
			\put(116,51){\rotatebox{90}{$\varepsilon_e(\rm MeV)$}}
			
		\end{picture}
	\setlength{\abovecaptionskip}{-0.7 cm}
	\caption{Distribution of spin polarization in $xy$-plane after a TSP (a) and (b) LSP electron beam passes through a DLW. (c) Number density of the electron beam (pcolor) in $xy$-plane and the corresponding beam profile (solid line) after exiting the DLW. The units are given in arbitrary units (a.u.). (d) Variation of normalized  number density with respect to time $t$ and electron energy $\varepsilon_e$ (pcolor), and temporal evolution of $\Delta E/E$ (dashed line). The THz intensity corresponds to $a_0=0.35$ with an initial duration of 5.7 ps, and initial electron energy is $\varepsilon_0=0.67$ MeV. The wavelength of incident THz wave is $\lambda_0 = 0.6 \rm ~mm$. The waveguide consists of a vacuum core with a radius of $0.7\lambda_0$ and a fused silica dielectric layer with a thickness of $0.4\lambda_0$.}
	\label{figspider}
\end{figure}

\begin{figure}[!t]
	\centering
		\includegraphics[width=1\linewidth]{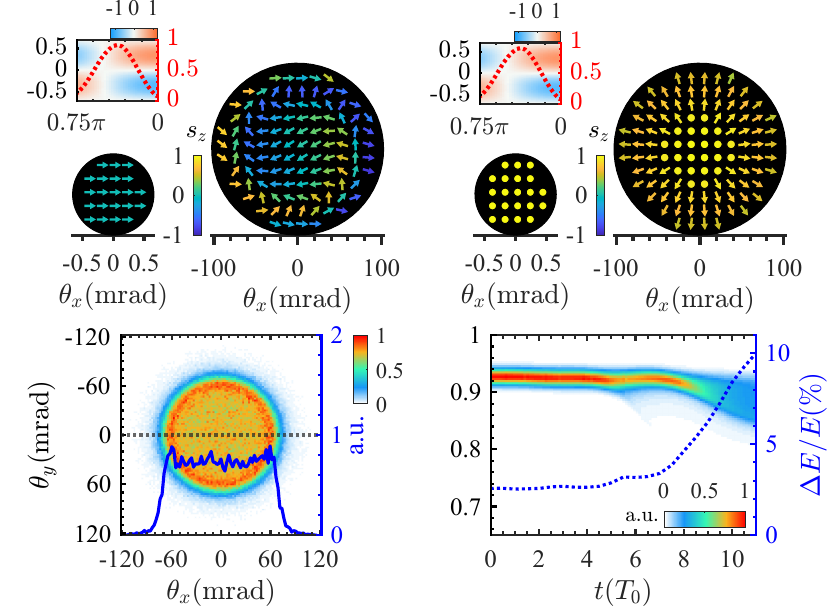}
		
		\begin{picture}(300,0) 
			\put(100,177){(a)}
			\put(24,172){\footnotesize (i)}
			\put(11,131){\footnotesize (ii)}
			\put(82,175){\footnotesize (iii)}
			\put(34,151){\footnotesize $\varphi$}
			\put(2,163){\footnotesize \rotatebox{90}{$y (\lambda_0)$}}
			\put(22,182){\footnotesize $E_x$}
			\put(60,163){\footnotesize \rotatebox{90}{\textcolor{red}{$|B_{z_0}|$}}}
			
			\put(220,177){(b)}
			\put(144,172){\footnotesize (i)}
			\put(130,131){\footnotesize (ii)}
			\put(202,175){\footnotesize (iii)}
			\put(154,151){\footnotesize $\varphi$}
			\put(123,162){\footnotesize \rotatebox{90}{$y (\lambda_0)$}}
			\put(142,182){\footnotesize $E_x$}
			\put(179,162){\footnotesize \rotatebox{90}{\textcolor{red}{$|B_{z_0}|$}}}
			
			\put(37,84){(c)}
			\put(104,95){\footnotesize a.u.}
			
			\put(147,84){(d)}
			\put(123,45){\rotatebox{90}{$\varepsilon_e(\rm MeV)$}}
			
		\end{picture}
		\setlength{\abovecaptionskip}{-0.7 cm}
	\caption{(a) Distribution of spin polarization in $(\theta_x,\theta_y)$ space after a TSP electron beam passes through a DLW. (a)-(i) represents variation of transverse electric field (pcolor) as a function of phase $\varphi=\omega t-kz+\varphi_0$ and $y$-coordinate, and variation of $|B_{z_0}|$ (dashed line) with respect to $\varphi$. $B_{z_0}$ is average longitudinal magnetic field sensed by the electrons. (a)-(ii) and (a)-(iii) are the initial and final spin-polarization distributions in $(\theta_x,\theta_y)$ space, respectively. (b) is analogous to (a) but for initial LSP electron beam. (c) Number density of the electron beam (pcolor) in $(\theta_x,\theta_y)$ space and the corresponding beam profile (solid line) after exiting the DLW. (d) Variation of normalized  number density with respect to $t$ and $\varepsilon_e$ (pcolor), and temporal evolution of $\Delta E/E$ (dashed line). The THz intensity corresponds to $a_0=1$ with an initial duration of 5.7 ps, and initial electron energy is $\varepsilon_0=0.93$~MeV. The wavelength of incident THz wave is $\lambda_0 = 0.6 \rm ~mm$. Vacuum core and dielectric layer thickness of the DLW are $0.7\lambda_0$ and $0.4\lambda_0$, respectively.}
	\label{fig2}
\end{figure}
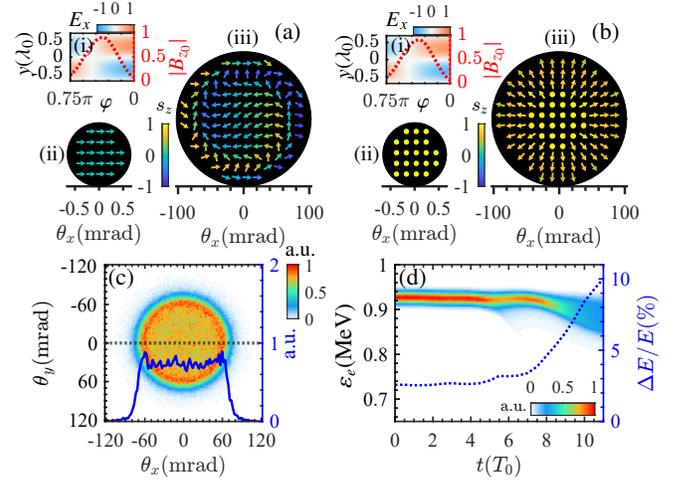

\begin{figure}[!t]
%
	
		\centering
        \flushright
	    \includegraphics[width=0.96\linewidth]{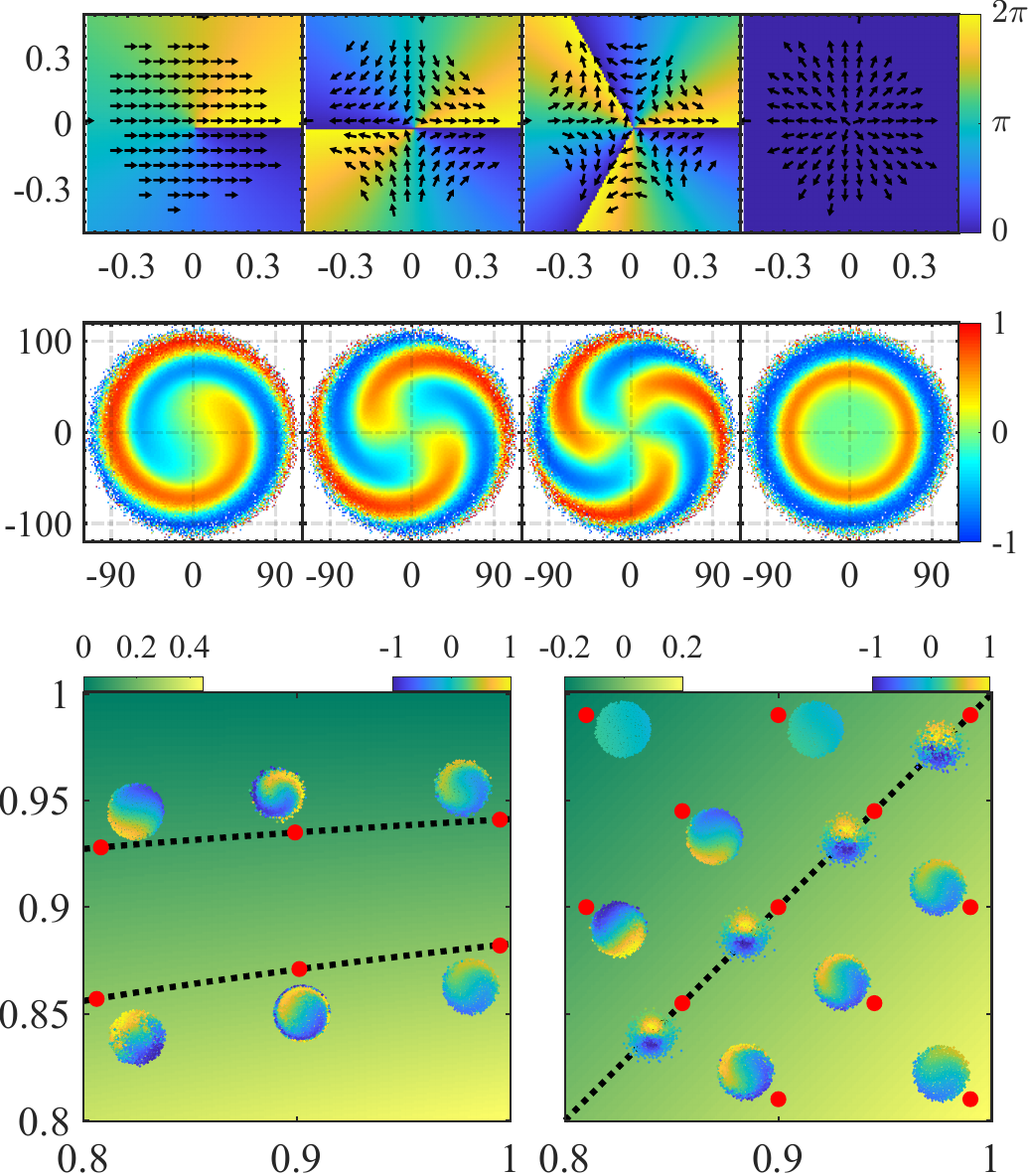}\vspace{30pt}
	    \begin{picture}(300,0) 
	    	\put(31,299){(a)}
	    	\put(48,239){$x (\lambda_0)$}
	    	\put(0,275){\rotatebox{90}{$y (\lambda_0)$}}
	    	\put(80,299){(b)}
	    	\put(98,239){$x (\lambda_0)$}
	    	\put(131,299){(c)}
	    	\put(148,239){$x (\lambda_0)$}
	    	\put(181,299){(d)}
	    	\put(198,239){$x (\lambda_0)$}
	    	\put(228,310){$\theta_B$}
	    	\put(31,228){(e)}
	    	\put(40,168){$\theta_x (\rm mrad)$}
	    	\put(0,195){\rotatebox{90}{$\theta_y (\rm mrad)$}}
	    	\put(80,228){(f)}
	    	\put(90,168){$\theta_x (\rm mrad)$}
	    	\put(131,228){(g)}
	    	\put(140,168){$\theta_x (\rm mrad)$}
	    	\put(181,228){(h)}
	    	\put(190,168){$\theta_x (\rm mrad)$}
	    	\put(228,240){$s_z$}
	    	
	    	\put(31,142){\textcolor{white}{(i)}}
	    	\put(0,100){\rotatebox{90}{$\beta_e$}}
	    	\put(68,33){ $v_p/c$}
	    	\put(57,152){ $\alpha$}
	    	\put(90,154){$s_z$}
	    	\put(36,87){1}
	    	\put(75,93){2}
	    	\put(115,95){3}
	    	
	    	\put(141,142){\textcolor{white}{(j)}}
	    	\put(180,33){$v_p/c$}
	        \put(172,154){\footnotesize {$\Delta\beta$}}
	    	\put(201,154){$s_z$}
	    	\put(193,55){1}
	    	\put(215,78){2}
	    	\put(230,89){3}
	    	
	    \end{picture}

	     \label{minipage1}
    \vspace{-20pt}
%
%
%

%
%
   \setlength{\abovecaptionskip}{-0.4 cm}
    \caption{(a)-(d) Distributions of $\theta_B$ (pcolor) and initial spin-polarization directions of incident electron beams (arrows) in the $xy$-plane when the incident mode is $\rm TE_{01}$. (e)-(h) Distributions of longitudinal spin polarization in the $(\theta_x,\theta_y)$ space. The spin-field azimuthal ratios $l_e$ are: (a) and (e) $l_e = 1$, (b) and (f) $l_e = 2$, (c) and (g) $l_e = 3$, and (d) and (h) $l_e = 0$. (i) and (j) are variations of $\alpha$ and $\Delta\beta$ with respect to $\beta_e$ and $v_p/c$. The dashed lines in (i) is the isolines of $\alpha$, indicating values of 0.12 and 0.25 for the upper and lower lines, respectively. Each inset image next to a red dot represents SP structure of electron beam in $(\theta_x,\theta_y)$ space at the corresponding $v_p/c$ and $\beta_e$ at that particular dot. }
    \label{fig3}

\end{figure}
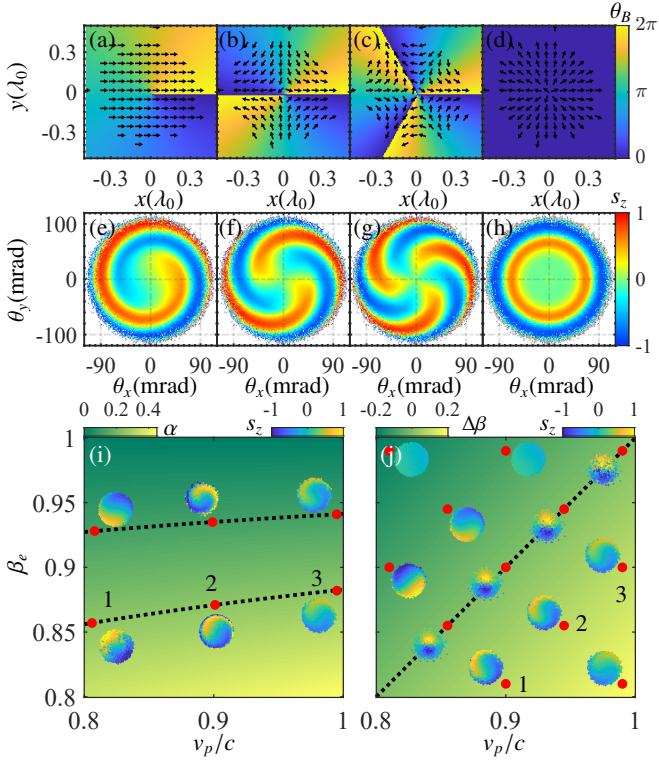

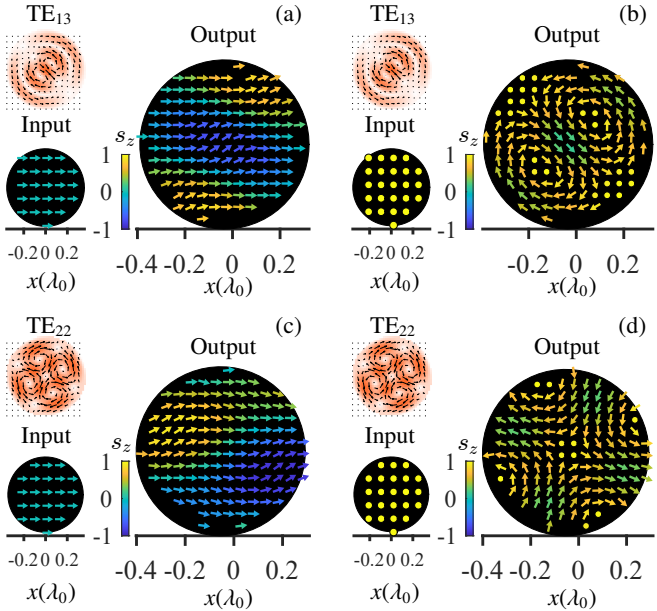
\begin{figure}[t] 
	\vspace{12pt}	
	\centering\includegraphics[width=1\linewidth]{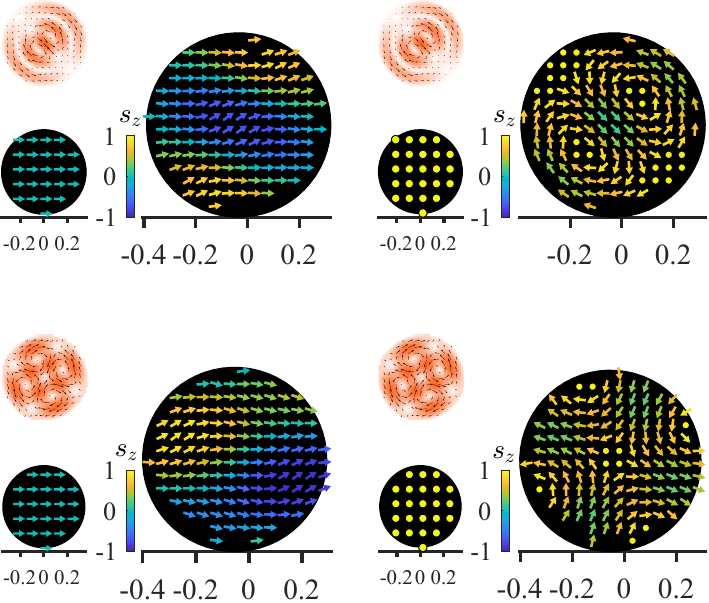}
	\begin{picture}(300,0)
		\put(100,223){ (a)}
		\put(8,223){$\rm TE_{13}$}
		\put(8,122){$x (\lambda_0)$}
		\put(75,120){$x (\lambda_0)$}
		\put(6,181){Input} 
		\put(70,215){Output}
		
		\put(230,223){ (b)}
		\put(138,223){$\rm TE_{13}$}
		\put(138,122){$x (\lambda_0)$}
		\put(205,120){$x (\lambda_0)$}
		\put(136,181){Input}
		\put(200,215){Output}
		
		\put(100,105){ (c)}
		\put(8,105){$\rm TE_{22}$}
		\put(8,4){$x (\lambda_0)$}
		\put(75,2){$x (\lambda_0)$}
		\put(6,63){Input}
		\put(70,97){Output}
		
		\put(232,105){(d)}
		\put(138,105){$\rm TE_{22}$}
		\put(138,4){$x (\lambda_0)$}
		\put(205,2){$x (\lambda_0)$}
		\put(136,63){Input}
		\put(200,97){Output}
		
	\end{picture}
	\setlength{\abovecaptionskip}{-0.4 cm}
	\setlength{\belowcaptionskip}{0.5 cm}
	\caption{Generation of SSP electron beams using high-order EM modes: (a)-(b) $\rm TE_{13}$, and (c)-(d) $\rm TE_{22}$, with initial polarization of electron beam being TSP for (a) and (c), and LSP for (b) and (d). In each panel, transverse distribution of spin polarization of a beam entering and exiting the DLW is displayed in the $xy-$plane. Pcolor and vectorgraph in upper left-hand corners represent intensity and direction of the transverse electric field, respectively.}
	\label{fig4}
\end{figure}

In this Letter, we propose a novel scheme for generating customizable SSP lepton beams by matching EM modes within the waveguide to spin-polarization modes of the lepton beams. For example, matching the $\rm TE_{01}$ mode \cite{Fisher22NPho} in a DLW with a transversely spin-polarized (TSP) lepton beam \cite{Hibberd20NPho} can produce a beam with a spider-like SP structure on the picosecond timescale, provided that the THz phase velocity $v_p$ is in proximity to the lepton velocity $v_l$; see Fig.~\ref{fig1}(a) and Fig.~\ref{figspider}. The essence of this mode matching arises from the spatially distributed nature of interaction between lepton spins and localized EM fields; see Fig.~\ref{fig1}(b). In addition, the difference between $v_l$ and $v_p$ results in a phase shift experienced by leptons, while also causing them to form a helically SP structure; see Fig.~\ref{fig1}(d) and Fig.~\ref{fig2}. We find that the angle between directions of the spin vector and the magnetic field $\theta_B$ is crucial for mode matching; see Fig. \ref{fig3}. Additionally, both the lepton velocity $v_l$ and the THz phase velocity $v_p$ also have significant impacts on the spin-polarization manipulation. Moreover, we demonstrate that customizable SP structures can be generated by matching high-order transverse electric (TE) modes with spin polarization states of the lepton beam; see Fig. \ref{fig4} and \cite{supp}. Our method is feasible with currently available THz sources and has capability to generate SSP lepton beams with customizable SP structures.

In our simulations, we set THz wave intensities within the range $0.1\lesssim a_0 \equiv eE_0/(m_e\omega c) \lesssim 1$ \cite{Thompson08PRL} and use electron beams as a representative example of leptons with energies at MeV level \cite{Wang89IEEE, Hassanein06PRL}. Here, $E_0$, $\omega$, $c$, $m_e$, and $e$ are THz electric field, angular frequency of the incident THz wave, speed of light in vacuum, and mass and charge of the electron, respectively. Our scheme is feasible, given that current THz pulse energies can reach joule level \cite{LGQ19PNAS, Bruhaug24OL} and peak powers can reach up to $\rm 1 ~TW$ \cite{LGQ20PRX}. Under these conditions, the quantum nonlinear parameter \cite{Gonoskov22RMP, Baier98, Ritus85JSLR} is calculated as $\chi=[e\hbar/(m_e^3 c^4)]\sqrt{-(F_{\mu\nu}p^\nu)^2}\sim10^{-8}\ll 1$, where $\hbar$, $F_{\mu\nu}$ and $p^\nu$ denote the reduced Planck's constant, field tensor and four-momentum of the particle, respectively. Consequently, the quantum radiation effects on the electron spin can be neglected. Besides, for the parameters of our simulations, the Stern-Gerlach and space-charge forces are much weaker than the Lorentz force \cite{Mane05RPP, Thomas20PRAB}. Thus, the spin dynamics are primarily governed by the Thomas-Bargmann-Michel-Telegdi equation \cite{1999Jackson}, while the momentum dynamics are governed by the relativistic Lorentz equation \cite{05Paul}. The pulse broadening caused by dispersion effects as the THz wave propagates through the DLW is taken into account \cite{Claude05, YEH78JOSA, supp, Naftaly21AS, 2005Microwave}. Moreover, longitudinal size of the electron beam is 28 fs, which is considerably smaller than the THz wavelength, and the propagation distance is relatively short. Therefore, all electrons experience nearly identical phase, and the wakefields resulting from the self generated EM fields of the beam reflecting off the DLW walls have negligible impacts on the electrons~\cite{Wong13OE, Kim10PRSTAB}. 

\textit{SSP electron beam---}When investigating mode matching mechanism within a DLW, we find that effective spatial manipulation of spin polarization can be achieved by using $\rm TE_{01}$ mode. This mode can be excited by an azimuthally polarized THz wave~\cite{Winnerl09OE, Ren06APL,Grosjean08OE}. When the THz phase velocity matches electron velocity, a spider-like or azimuthally SP structure can be generated depending on the initial spin-polarization states of the electron beams, as shown in Fig.~\ref{figspider}. The initial phase of the electron beam is matched to 0 during its interaction with the THz wave, and the interaction length can be derived as $L=\tau v_g v_e/(v_e-v_g)$, where $v_e$, $\tau$ and $v_g$, are electron velocity, duration and group velocity of the THz pulse, respectively. Due to proximity of the electron velocity to the THz phase velocity ($v_p \approx v_e$), the EM fields within the DLW exert negligible influence on the transverse momentum of electrons. Thus, the transverse size remains stable, and spin polarization of the beam is shown in the $xy$-plane in Fig.~\ref{figspider}. Besides, the energy spread is reduced from 2.6\% to 1.4\%, while the average energy exhibits no significant change.

We also find that, even when the velocities are not matched, it is still possible to generate SSP electron beams and maintain the beam quality. This is because that the $\rm TE_{01}$ mode exhibits central symmetry, and the suitable longitudinal magnetic field serves to partially suppress the divergence of the electron beam. As shown in Figs. \ref{fig2}(a)-(b), the initial TSP electron beam transforms into a helically SP structure in $(\theta_x,\theta_y)$ space after passing through the DLW, while the initial longitudinally spin-polarized (LSP) electron beam exhibits a radially SP structure. During the manipulation, the initial phase experienced by electrons is $\varphi_i=0.75\pi$, and the final phase is $\varphi_f=0$. Phase jitter within $0.1\pi$ can be considered to impact spin manipulation negligibly, and further information regarding the SP structures for various phase ranges are provided in \cite{supp}. 
Typically, an electron is subjected to transverse Lorentz forces in the DLW and undergoes helical motion, with a sense of gyration (clockwise or counterclockwise relative to the propagation direction) determined by direction of the longitudinal magnetic field \cite{supp}. Density of the electrons exiting the DLW in $(\theta_x,\theta_y)$ space deviates from the initial Gaussian distribution, evolving into a flattened-top shape, as shown in Fig.~\ref{fig2}(c). The rms divergence of the electron beam in Fig. \ref{fig2} are 0.17 mrad before entering the DLW and 54.44 mrad upon exiting. Besides, the energy spread increases from 2.6\% to 10\%, and the average energy decreases by about 4\%. These changes are attributed to the transverse Lorentz force on electrons with velocity $v_e$. 

\textit{Spatial manipulation of spin polarization---}This is associated with both $\bm{\Omega}_{\parallel}$ and $\theta_B$. Here, $\bm{\Omega}_{\parallel}=-(e/m_ec) \alpha B_\perp$ represents the precession frequency of the longitudinal spin component, while $\theta_B$ denotes the angle between directions of the spin vector and the magnetic field. Notably, the spatial distribution of $\theta_B$ is crucial for spin manipulation, as it reflects the condition of mode matching.
In this context, the longitudinal precession frequency coefficient $\alpha = \left(g/2-1\right)-\left(g\beta_e/2-1/\beta_e\right)v_p/c \in (0,\infty)$ characterizes the effect of electron velocity and THz phase velocity on the precession frequency. The terms $\beta_e$, $B_\perp$, $g$, and $\bm{s}$ are the electron velocity normalized to $c$, amplitude of the transverse magnetic field, Land\'e $g-$factor, and electron spin, respectively. Consequently, the rate of change of longitudinal polarization can be written as
$\left(d/dt\right)\left(\hat{\bm{\beta_e}}\cdot\bm{s}\right) =\bm{\Omega}_{\parallel}s_\perp \sin(\theta_B)$ \cite{1999Jackson,supp}, where $\hat{\bm{\beta_e}}$ is a unit vector in the direction of $\bm{\beta_e}$, and $s_\perp$ represents the spin component perpendicular to the electron velocity. The precession frequency, which is related to the magnetic field strength, electron velocity, and THz phase velocity, determines the efficiency of spin manipulation of the electron beam. When an electron beam propagates within a DLW, all electrons experiencing the same field intensity exhibit the same precession frequency $\bm{\Omega}_{\parallel}$. This implies that, regardless of $\theta_B$, spin polarization of the electrons simultaneously reaches maximum or minimum values during their precession. However, distribution of extreme values in the longitudinal polarization of electron beam is related to $\theta_B$, which has a spatial distribution that corresponds to the matching between spin-polarization and EM mode. Taking the helically SP structure as an example, it is clear that the dependence of $\theta_B$ on the azimuthal angle at a constant radius is analogous to the transverse phase dependence of vortex light beams, as shown in Figs.~\ref{fig3}(a)-(d). The number of spiral stripes depends on the spin-field azimuthal ratio $l_e=\theta_B / \phi$, which describes the relationship between $\theta_B$ and the azimuthal angle $\phi$, as shown in Figs.~\ref{fig3}(a)-(h). Base on this, we can achieve control over the number of spiral stripes. Furthermore, when $l_e$ is negative, the chirality of the spiral stripes is reversed. For example, in the cases where $l_e=-1$ and $l_e=1$, the number of stripes is the same, but the chirality is opposite.

Note that the efficiency of spin manipulation primarily depends on the precession frequency, which in turn depends on the intensity of incident THz wave and $\alpha$. The THz intensity depends on the THz technology and the damage threshold of the DLW, while $\alpha$ can be altered by controlling the THz phase velocity and electron velocity. We investigate dependence of $\alpha$ on $v_p/c$ and $\beta_e$; see Fig. \ref{fig3}(i). Overall, $\alpha$ is more sensitive to $\beta_e$ than to $v_p/c$. For the same magnetic field, a smaller $\beta_e$ and a larger $v_p/c$ result in a higher value of $\alpha$, thereby increasing the efficiency of spin manipulation. Due to the mismatch between $\beta_e$ and $v_p/c$, the electrons undergo a phase shift when interacting with THz waves. To ensure that the electrons experience a consistent phase range, we maintain a THz phase window of $(0.5\pi,\pi)$, which results in a phase experienced by the electrons from $\pi$ to $0.5\pi$, for $\beta_e>v_p/c$, and $0.5\pi$ to $\pi$, for $v_p/c>\beta_e$. Within the same phase range, the interaction time between electrons and THz waves is inversely proportional to $|\Delta\beta|$ \cite{supp}. Here, $\Delta\beta=v_p/c-\beta_e$ represents the difference between normalized electron velocity and THz phase velocity, indicating the transverse Lorentz force experienced by the electron and signifying the interaction time. As $v_p/c$ and $\beta_e$ approach the line $v_p/c=\beta_e$, the interaction time increases, with the same phase range. Hence, it is necessary to consider the combination of $\alpha$ and $\Delta \beta$. On the one hand, it is essential to maximize $\alpha$ to ensure an adequate spin-manipulation efficiency. On the other hand, one needs to increase the interaction time and minimize the transverse Lorentz force, to enhance visibility of the SP structure and preserve the electron-beam quality. It is shown in Fig. \ref{fig3}(i) that, even for the same $\alpha$, visibility of the SP structure is different, since the total interaction time is different, e.g., points 1, 2 and 3 in Fig. \ref{fig3}(i). When $\beta_e$ and $v_p/c$ are parallel to the line $v_p/c=\beta_e$, e.g., points 1, 2 and 3 in Fig.~\ref{fig3}(j), the electron dynamics and total interaction time are the same. In this situation, visibility of the SP structure is positively correlated with $\alpha$.

\textit{Spin-polarization structures using high-order TE modes---}As shown in Fig. \ref{fig4}, employing high-order TE modes, the resulting SP structure of the electron beam is totally different from that produced in a $\rm TE_{01}$ mode. In order to mitigate the influence of high-order modes on the transverse
distribution of electron beams, we make $\beta_e\approx v_p/c$. From Fig. \ref{fig4}, it can be inferred that by controlling the modes matching condition within the DLW, it is possible to achieve ultrafast manipulation of SP structures with customizable configurations. For SP structures under some other EM modes; see \cite{supp}. In addition, LSP electron beams may act as a mode detector in the DLW, as direction of transverse spin polarization of the electron beam is the same as that of the electric field locally; see Figs. \ref{fig4}(b) and (d). This provides a new diagnostic measure for EM modes inside DLWs.

In conclusion, we have put forward a novel and flexible method of ultrafast generation of relativistic SSP lepton beams using mode matching within a DLW. Relativistic SSP lepton beams have the potential to generate high-energy structured photon beams and pave the way for novel research avenues in relativistic structured particle beams, especially in nuclear physics, high-energy physics, materials science and atomic physics.


\bigskip

{\it Acknowledgement:} The work is supported by the National Natural Science Foundation of China (Grants No. 12425510, No. U2267204, No. 12441506, No. 12405171, No. 12275209, No. 12475249, No. 12447106), the National Key Research and Development (R\&D) Program (Grants No. 2024YFA1610900, No. 2024YFA1612700), Natural Science Basic Research Program of Shaanxi (Grant No. 2024JC-YBQN-0042), the Fundamental Research Funds for Central Universities (No. xzy012023046), and the Innovative Scientific Program of CNNC.  Y. I. S. is supported by an American University of Sharjah Faculty Research Grant (FRG23-E-S81) and acknowledges hospitality at the School of Physics, Xi'an Jiaotong University.


\bibliography{refs}

\end{document}